\documentclass[aps,showpacs,twocolumn, tightenlines]{revtex4}
\usepackage{epsfig,amssymb,bm,graphics}

\begin{document}


\title{
Coherent  $\bm \phi$ and $\bm \omega$  meson photoproduction from
deuteron and non-diffractive channels}


 \author{A.I.~Titov$^{ab}$, M. Fujiwara$^{ac}$, and T.S.-H. Lee$^{d}$}
 \affiliation{
 $^a$Advanced Science Research Center, Japan Atomic Energy Research Institute,
 Tokai, Ibaraki,
 319-1195, Japan\\
 $^b$Bogoliubov Laboratory of Theoretical Physics, JINR,
  Dubna 141980, Russia\\
 $^c$Research Center of Nuclear Physics, Osaka University,
  Ibaraki, Osaka 567-0047, Japan\\
 $^d$Physics Division, Argonne National Laboratory, Argonne,
  Illinois 60439, USA}


\begin{abstract}
 For coherent photoproduction of $\phi$ and $\omega$ mesons
 from deuteron at forward angles,
 the isovector $\pi$-exchange amplitude is strongly suppressed.
 We show a possibility to study
 the non-diffractive channels associated with the unnatural
 parity exchange in $\phi$-photoproduction and with the baryon resonance
 excitations in $\omega$-photoproduction by
 measuring the spin observables.
\end{abstract}


 \maketitle
  The prime interest in the $\phi$ and $\omega$ meson photoproduction at
  a few GeV is related to the possible manifestation of
  non-diffractive "exotic" channels.
  In $\phi$ photoproduction, the amplitude of the  unnatural
  parity-exchange processes is responsible for
  such exotic mechanisms  as the direct
  knockout of $\bar ss$ component of the strangeness sea in a nucleon
  (cf.~\cite{Ellis99,Henley92,TOYM98} for references) and the anomaly
  Regge trajectories associated with the
  non-perturbative gluon-exchange processes~\cite{Kochelev01}.
  In the  $\omega$ photoproduction it is the baryon resonance
  excitation~\cite{ZLB98,OTL01,TL02}, which is closely related
  to several aspects
  of intermediate and
  high  energy physics, ranging
  from resolving the so-called  "missing" resonance
  problem~\cite{IK77-80} to estimating in-medium modifications
  on the vector meson
  properties~\cite{FP97}.

  The contribution of the exotic non-diffractive channels
  to the total unpolarized cross section is rather small
  compared to the dominant
  diffractive Pomeron exchange  in $\phi$ photoproduction and
  the $\pi$-meson exchange in $\omega$ photoproduction.
  Therefore, we are pinning our hopes on the measurement of  spin observables.
  These could select the amplitudes with different parity-exchange
  symmetries, like the decay asymmetry $\Sigma_V$,
   ($V=\omega,\phi$)~\cite{Tabakin,TLTS99},
  or to  find such observables which are  proportional to the
  interference terms of
  the amplitudes with different parity properties, like
  beam-target asymmetry  in reactions with circular polarized
  beam and a polarized target~\cite{TOYM98,Tabakin}.
  The central problem when studying spin observables is the
  relatively strong
  influence of the unnatural parity  $\pi$-exchange
  amplitude (PEA). In $\omega$ photoproduction, PEA dominates more than
  90\% of the resonance excitation.
  In the $\phi$ photoproduction at forward angles, the
  PEA contribution is comparable to those expected from
  exotic non-diffractive channels. This situation causes considerable
  difficulties in extracting the true
  exotic-channel contributions from
  the forthcoming experimental results from LEPS at SPring-8,
  Thomas Jefferson National Accelerator Facility,
  ELSA-SAPHIR at Bonn, and  GRAAL at Grenoble.

  One of the possible solutions to eliminate the contribution
  of the isovector $\pi$-meson exchange process
  is to use an isoscalar target.
  The simplest case is coherent photoproduction from the deuteron.
  The deuteron, with spin 1, has an advantage compared to
  spinless isoscalar targets. It gives the opportunity
  to examine the various beam-target asymmetries, which are
  sensitive to the non-diffractive channels.

  In the present Rapid Communication, we wish to report this particular
  important and interesting aspect of the coherent  $\phi$ and $\omega$
  photoproduction from deuteron.
  The coherent photoproduction from the deuteron at higher
  energies with  different points of view and  purpose
  was analyzed in Ref.~\cite{FKS97}.

  We define the kinematical variables for the $\gamma \to VD$
  reaction  with  usual notations.
  The four momenta of the
  incoming photon, outgoing vector meson, initial deuteron, and final
  deuteron are denoted as $k$, $k_V$, $p$ and $p'$,
   respectively. The standard Mandelstam variables are
   defined as  $t = (p' - p)^2
   = (k-k_V)^2=-Q^2$, $s \equiv W^2 = (p+k)^2$.
   The space component of the transferred momentum transfers to deuteron in
   the laboratory system is ${\bf q}^2\equiv q^2=Q^2(1+Q^2/4M_D^2)$, where
   $M_D$ is the deuteron mass.

    We consider the case where the initial photon energy is below
    $2-3$ GeV and momentum transfer $Q^2$ is smaller than $0.5$ GeV$^2$.
    Under these conditions, the dominant contribution to the amplitude
    comes from the
    single scattering processes, which are depicted in
    \begin{figure}[ht]
   \includegraphics[width=0.7\columnwidth]
   {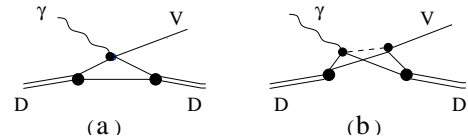}
   \caption{Diagrammatic  representation of
   (a) single and (b) double scattering
   contributions in the $\gamma D\to VD$ reaction.}
   \label{Fig1}
   \end{figure}
    Fig.~1a.  The double scattering
    diagrams shown in Fig.~1b are suppressed at low $Q^2$~\cite{FKS97}.
    At considered momentum transfers
    a non-relativistic approach for the deuteron structure
    is quite reasonable.  We use
    the deuteron wave function with a realistic (Paris) nucleon-nucleon
    potential~\cite{Paris}, which fairly well reproduces
    the deuteron electromagnetic form factor
    for $Q^2\leq 1$ GeV$^2$~\cite{ParisD} and has been used
    successfully in describing the $\eta$ photoproduction~\cite{Friman_eta}.

\begin{figure}[hb]
 \includegraphics[width=0.7\columnwidth]{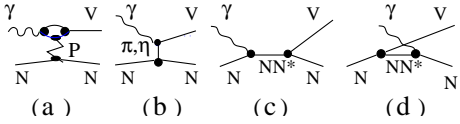}
 \caption{Diagrammatic representation of vector meson photoproduction from nucleon:
 (a) Pomeron exchange contribution, (b) pseudoscalar $\pi$ and
 $\eta$ exchange and (c) direct and (d) crossed $N$ and $N^*$ exchanged
 processes.}
 \label{Fig2}
 \end{figure}

 The total vector meson photoproduction amplitude in the reaction $\gamma
 D\to VD$ reads
 \begin{eqnarray}
 T^D_{M_fM_i;\lambda_V\lambda_\gamma}= 2 \sum_{\alpha\beta}
 \langle M_f\lambda_V,\beta|
 T^s_{\beta\alpha;\,\lambda_V\lambda_\gamma}
 |M_i\lambda_\gamma,\alpha\rangle,
\label{A1}
 \end{eqnarray}
 where $M_i,M_f,\lambda_\gamma$, and $\lambda_V$ stand for the  deuteron-spin
 projections of the initial and the final states, and helicities of
 the incoming photon and the outgoing vector meson, respectively.
 $T^s$ is the amplitude of the vector meson
 photoproduction from the isoscalar nucleon
 \begin{eqnarray}
 T^s\equiv\frac12(T^p + T^n ).
 \label{A1_2}
 \end{eqnarray}
 The indices $\alpha$
 and $\beta$ in Eq.~\ref{A1} refer to all quantum numbers before and after
 the collision. For the "elementary" photoproduction amplitudes
 $T^{p,n}$,
 we use the Pomeron-exchange contribution, pseudoscalar $\pi$ and
 $\eta$ exchange and direct and crossed $N$ and $N^*$ exchanged
 amplitudes shown in Fig.~\ref{Fig2}
 and described in~Refs.~\cite{TLTS99,TL02}.
 Because of Eq.~\ref{A1_2} the isovector
 $\pi$-exchange terms in the total amplitude are cancelled since
 $T^n_\pi=-T^p_\pi$.

  Using the standard decomposition of the deuteron state
  in terms of $s$ $(U_0)$ and $d$ $(U_2)$ wave functions,
   one can rewrite Eq.~\ref{A1} in the explicit form
 \begin{eqnarray}
&&  T^D_{M_f,M_i;\lambda_V\lambda_\gamma}(t)=2\sqrt{4\pi}\sum
  i^\lambda\,\frac{\widehat{L'}\widehat{\lambda}}{\widehat{L}}
  Y_{\lambda\mu}(\widehat{\bf q})\nonumber\\
&&
  C_{{\textstyle \frac12} m_1{\textstyle \frac12} m}^{1M}
  C_{{\textstyle \frac12} m_1'{\textstyle \frac12} m}^{1M'}
  C_{1M LM_L}^{1M_i}
  C_{1M' L'M_{L'}}^{1M_f}\nonumber\\
&&C_{L'M_{L'} \lambda\mu}^{LM_{L}}
  C_{L'0 \lambda  0}^{L0}
  \,R_{LL'\lambda}({q^2})\,
   T^s_{m_1m_1';\lambda_V\lambda_\gamma}(t),
   \label{A2}
 \end{eqnarray}
where $\widehat{j}=\sqrt{2j+1}$, and the radial  integral
$R_{LL'\lambda}$ reads
\begin{eqnarray}
  R_{LL'\lambda}(q^2)= \int dr
  U_L(r)U_{L'}(r)j_\lambda({qr}/{2}).
   \label{A3}
 \end{eqnarray}
Eq.~\ref{A2} is simplified if one chooses the quantization axis
along the transferred momentum ${\bf q}$ and only keeps  the
spin/helicity conserving terms with the natural $T^{\rm N}$ and
unnatural $T^{\rm U}$ parity exchange in the total amplitude
\begin{eqnarray}
  T^{{\rm N}\atop{\rm U}}_{m_1m_1';\lambda_V\lambda_\gamma}(t)=
  \left({{1}\atop{2m_1\lambda_\gamma}}\right)
  \delta_{m_1m_1'}\delta_{\lambda_\gamma\lambda_V}
  T^{{\rm N}\atop{\rm U}}_0(t).
   \label{A4}
 \end{eqnarray}
 Here, $T^{{\rm N}\atop{\rm U}}_0(t)$ is the scalar, spin-independent part of the
 amplitudes. Using
 Eq.~\ref{A2} with Eq.~\ref{A4}, we get the following result for the
 natural and unnatural parity-exchange parts of the total  amplitude
\begin{eqnarray}
  &&T^{D{\rm N}}_{M_fM_i;\lambda_V\lambda_\gamma} =
  2\delta_{M_iM_f}\delta_{\lambda_\gamma\lambda_V}
  (\delta_{\pm1M_i}S_1^{\rm N}  
   +\delta_{0M_i}S_0^{\rm
  N})
   T^{{\rm N}}_0,\nonumber\\
&& T^{D{\rm U}}_{M_fM_i;\lambda_V\lambda_\gamma}=
   2M_i\lambda_\gamma\delta_{M_iM_f}\delta_{\lambda_\gamma\lambda_V}
  \delta_{\pm1M_i}\,S_1^{\rm U}\,
   T^{{\rm U}}_0.
   \label{A5}
 \end{eqnarray}
 The form factors $S_M^{N,U}$ are similar to
 the deuteron electromagnetic form factors. The form factors for the
 natural exchange amplitude $S^{N}_{1,0}$
 corresponds to the electric form factors, and  are expressed
 as the combination of  charge $( F_C=R_{000} + R_{220})$
 and quadrupole
 $(F_Q =R_{202} - R_{222}/{\sqrt{8}})$ form factors.
  The form factor for the unnatural parity exchange amplitude
  $S^U_1$ is equal
  to the  magnetic form factor $F_M=R_{000} - R_{220}/2 +
 \sqrt{2} R_{202}+  R_{222}$. We thus write
 \begin{eqnarray}
 S^{N}_1(q^2)&=& F_C(q^2) - \frac{1}{\sqrt{2}}
 F_Q(q^2),\,\, 
 S^{U}_1(q^2)= F_M(q^2),\nonumber\\
 S^{N}_0(q^2)&=& F_C(q^2) + \sqrt{2} F_Q(q^2). 
 \label{A6}
 \end{eqnarray}
  Note that the presentation of the total amplitude in Eq.~\ref{A5}
  is valid at extremely small
  transferred momentum,  when one can neglect spin-flip terms in the
  elementary amplitudes. Also it assumes the quantization axis is along
  ($\widehat{\bf q}$), however for some spin observables
   it is chosen
  along the beam velocity in center of mass or in
  the vector meson rest frame (Gottried-Jackson system), which in general, is
  different from the direction of ($\widehat{\bf q}$).
 These conditions are realized at forward $\phi$-photoproduction where the
 possible spin-flip diagrams in the $N$ and $N^*$ exchange processes  are
 suppressed. In the case of the finite transferred momentum
 these simplification do not hold and one
 has to take into
 account all these effects and have to use the exact form in Eq.~\ref{A2},
 which we have done in our numerical calculations.
 But, Eq.~\ref{A5}  allows us
 to understand clearly the underlying physical meaning of photoproduction from deuteron, and
 therefore we can use it for the qualitative analysis.

\begin{figure}[h]
 \includegraphics[width=0.55\columnwidth]{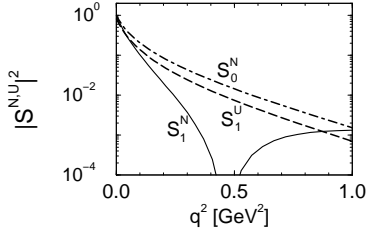}
 \caption{Deuteron form factors $|S^{\rm N,U}|^2$.}
 \label{Fig3}
 \end{figure}

 Let us start
 from the $\phi$ photoproduction.
 Inspection of Eqs.~\ref{A5} and \ref{A6} results in
 the following:\\
1. $T^D$ decreases with $-t$ much faster then elementary amplitude
$T^s$, because
  the form factors $S^{N,U}$ decrease rapidly.\\
2. The elementary spin conserving amplitude $T^s$ generates the
spin
  conserving $T^D$-amplitude.\\
3. Unnatural parity-exchange transitions are suppressed for the
deuteron target
  with spin
  projection $M_i=0$.\\
4. The form factors of the natural parity-exchange (Pomeron)
amplitudes with
  spin polarization $M_{i,f}=\pm 1$ and $M_{i,f}=0$ are different.
  Moreover, both
  of them are different from the unnatural parity-exchange form factor $S^U_1$:
   $S^N_{1}$ decreases much faster  with $q^2$  than the form factors
   $S^N_0$.
   The form factor $S^N_{1}$ has a node at $q^2\simeq 0.5$ GeV$^2$.
   This difference is illustrated in Fig.~\ref{Fig3}, where
   we show the $q^2$ - dependence of $|S^{N,U}|^2$.\\
5.  Contribution of the isovector $\pi$-exchange amplitude is
  strongly suppressed.\\
 Item (1) is illustrated in Fig.~4, where the unpolarized
 differential cross sections for the $\gamma p\to\phi p$
 and $\gamma D\to\phi D$ for $E_\gamma=2.2$ GeV
 reactions are presented.
 \begin{figure}[h]
 \includegraphics[width=0.45\columnwidth]{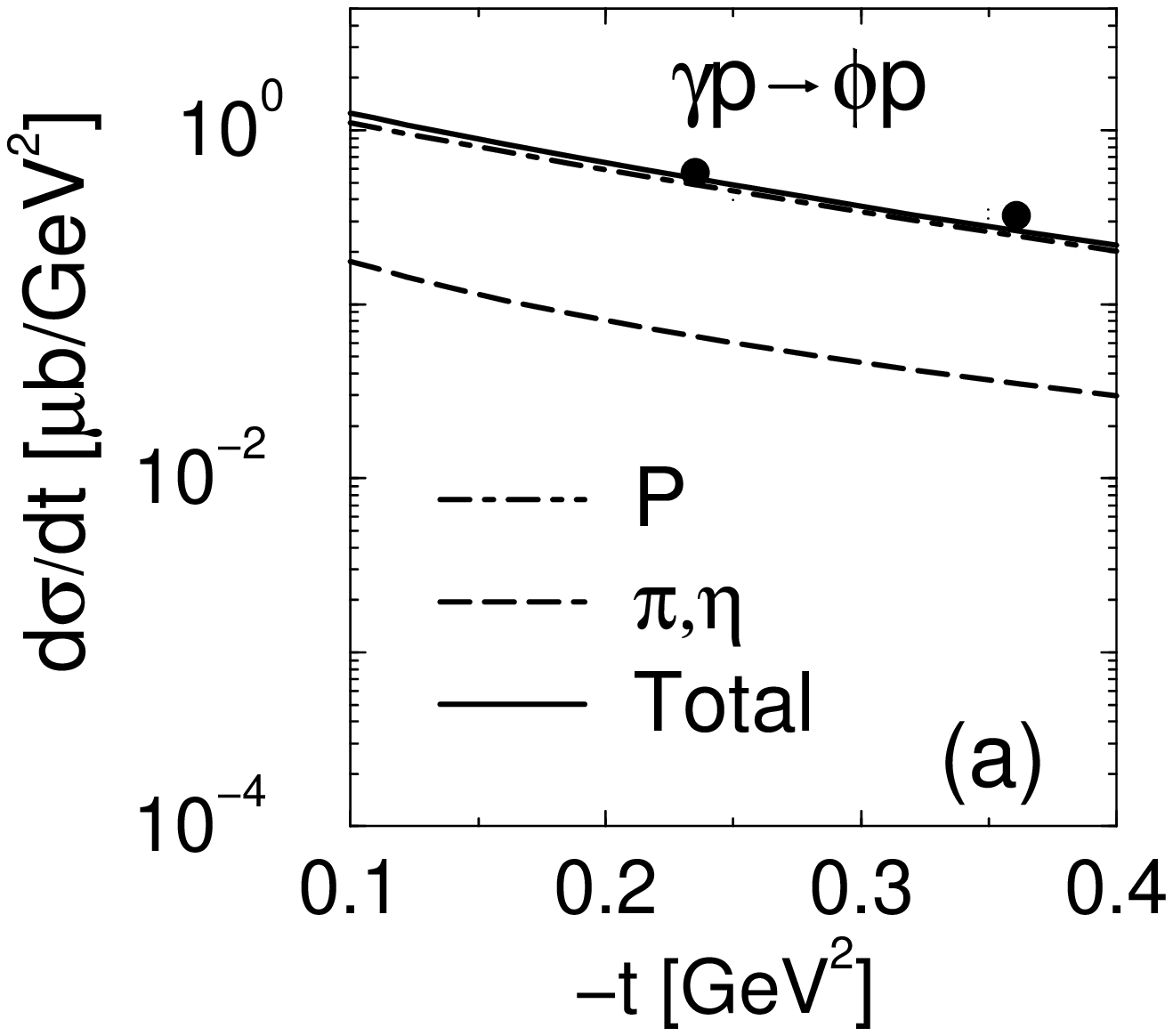}
 \includegraphics[width=0.45\columnwidth]{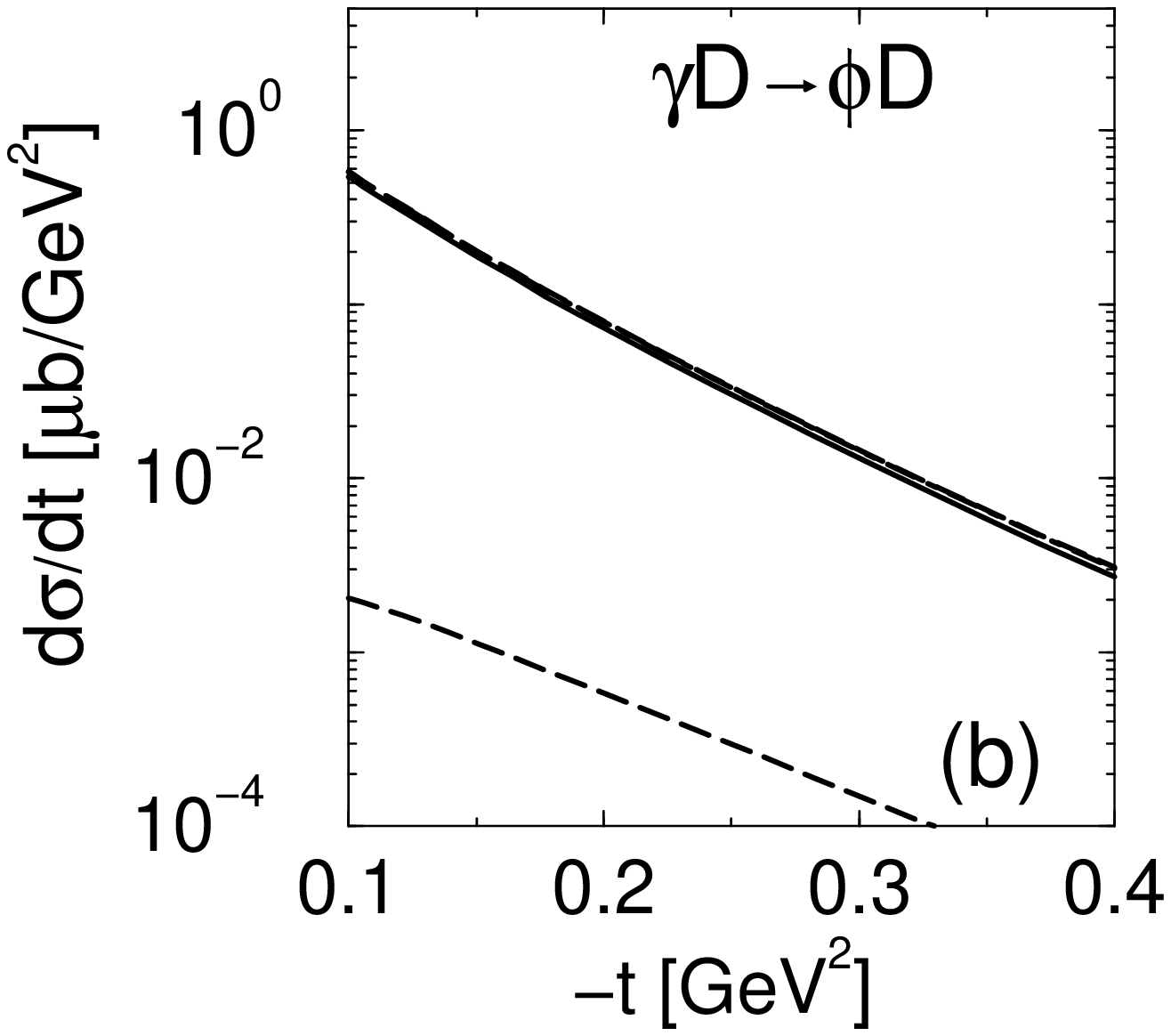}
 \caption{Differential cross sections of the $\gamma p\to \phi p$
 (a) and $\gamma D\to \phi D$ (b) reactions. Only the dominant
 contribution from the Pomeron,
 $\pi,\eta$ meson exchange processes are shown.
  Data are taken from Ref. \protect\cite{Bonn}}
 \label{Fig4}
 \end{figure}
 One can see that
 the slope of ${d\sigma^{\gamma D}}/{dt}$ is steeper  than that
 of ${d\sigma^{\gamma p}}/{dt}$ and the ratio of these cross sections at
 $|t|\sim |t_{\rm max}|\simeq 0.1$ GeV$^2$ is determined
 by the form factors $|S^N(t_{\rm max})|^2 < 1$
 (cf. Fig.~\ref{Fig3}).  The relative contribution of the unnatural-parity
 exchange (pseudoscalar-exchange) channel in the $\gamma D\to\phi D$
 reaction is much smaller than those in the $\gamma p\to\phi p$ reaction, because
 it comes only  from the $\eta$-meson exchange which is much
 smaller than the contribution of $\pi$-meson exchange in the $\gamma p\to\phi p$
 reaction~\cite{TLTS99}.

\begin{figure}[h]
 \includegraphics[width=0.43\columnwidth]{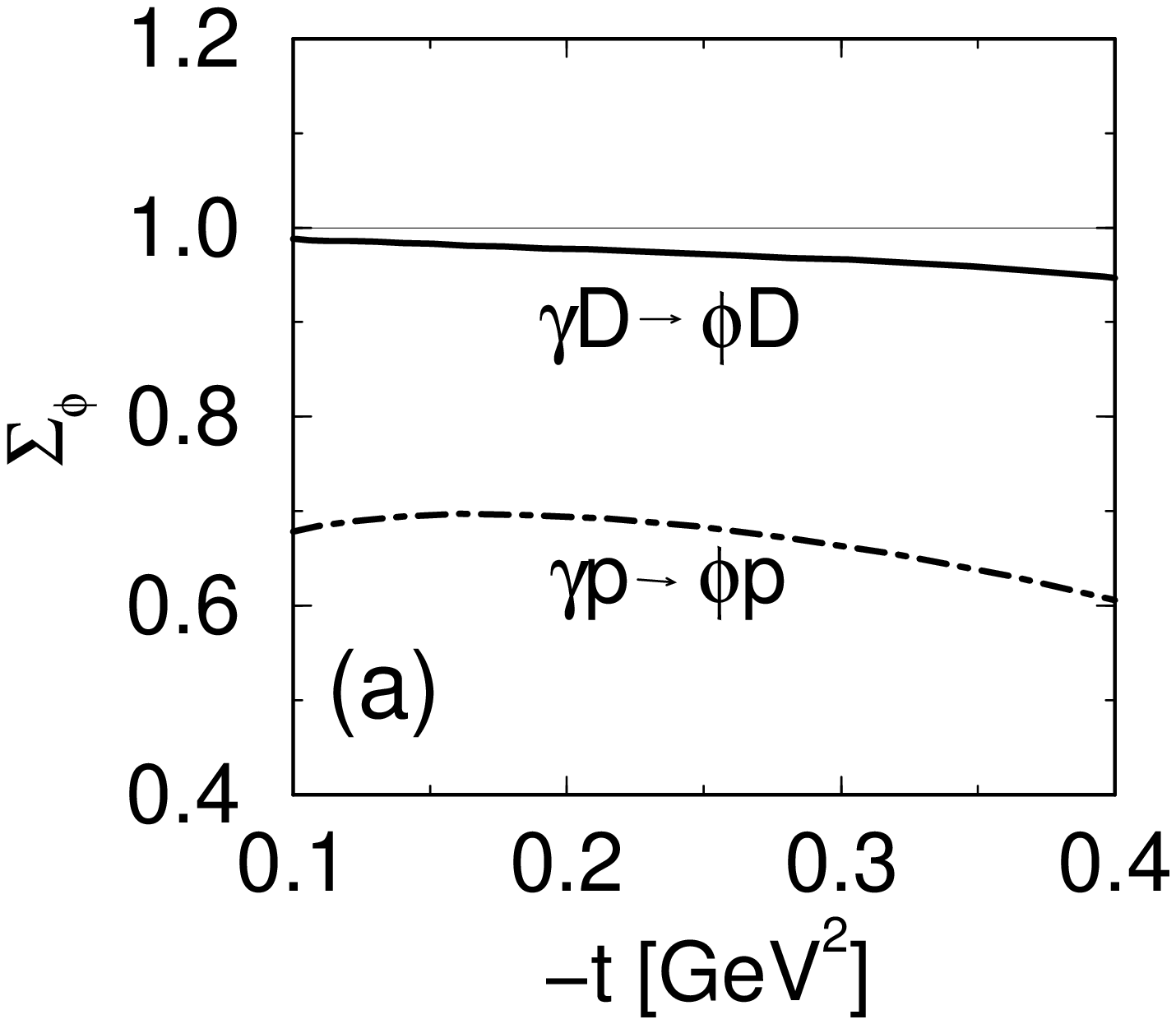}\qquad
 \includegraphics[width=0.43\columnwidth]{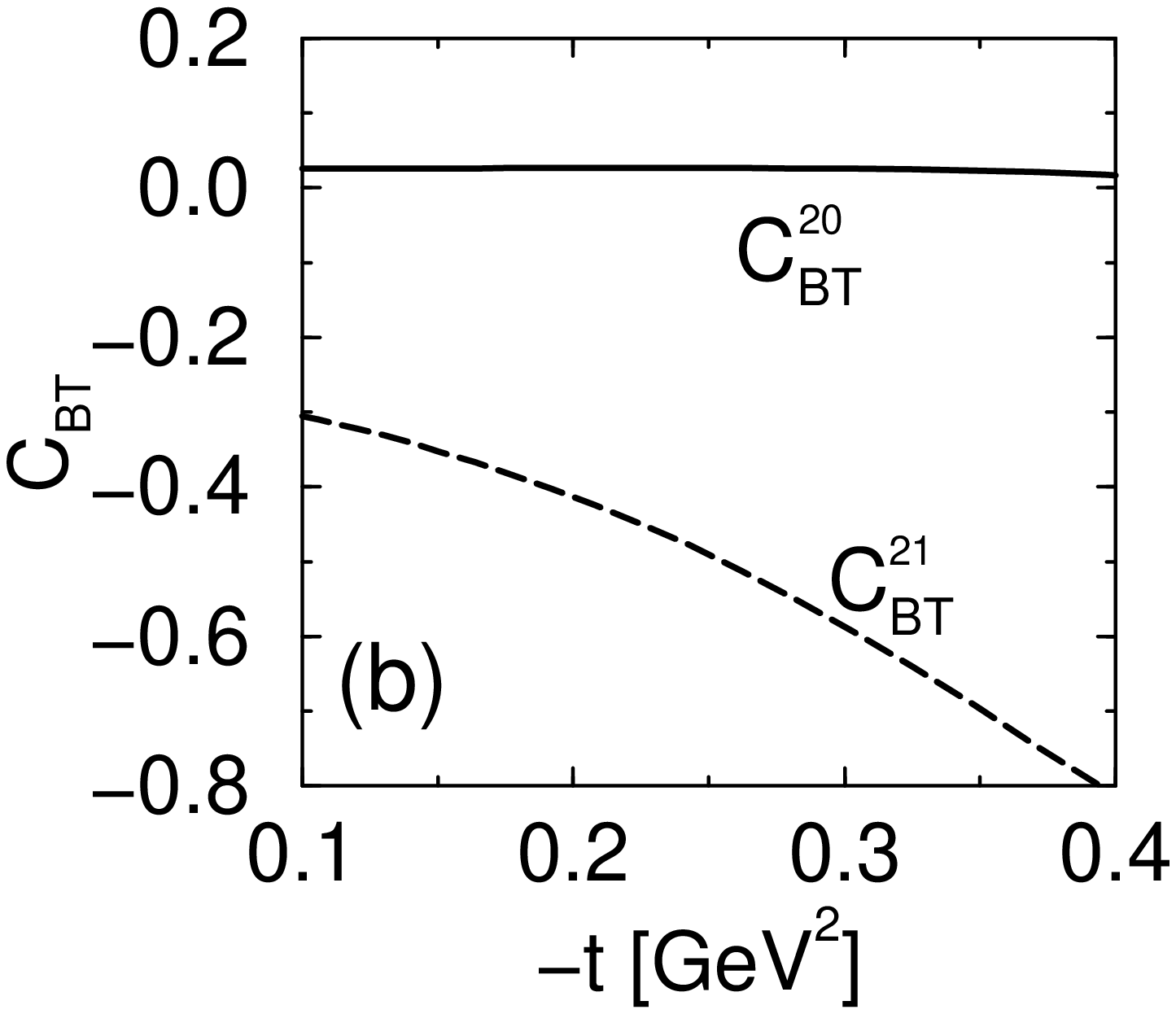}
 \caption{(a): The $\phi$ meson decay  asymmetries  for the
  $\gamma p\to \phi p$
  and $\gamma D\to \phi D$ reactions.
  (b): Beam-target asymmetries $ C^{21}_{BT}$ and $ C^{20}_{BT}$
  for the $\gamma D\to \phi D$ reaction. }
 \label{Fig5}
 \end{figure}

 The other items 2 to 5 mentioned above,
 are important   for the spin observables.
 Let us consider the $\phi$-meson decay asymmetry $\Sigma_{V=\phi}$,
 defined through the spin-density
 matrix $\rho^{i}_{\lambda\lambda'}$~\cite{TLTS99}
\begin{eqnarray}
 \Sigma_V=\frac{\rho^{1}_{11} + \rho^{1}_{1-1}}
  {\rho^{0}_{11} + \rho^{0}_{1-1}}.
\label{A8}
 \end{eqnarray}
 It has a definite value $\Sigma_V=+1(-1)$ for the natural (unnatural) parity-exchange
 amplitude, taken separately. Using this notation and Eq.~\ref{A5} we can estimate
 $\Sigma^p_\phi$ and  $\Sigma^D_\phi$ for $\gamma p$ and $\gamma D$ reactions,
 respectively, and investigate them simultaneously at $t\simeq t_{\rm max}$
 \begin{eqnarray}
 \Sigma^p_\phi &\simeq&1-2|\alpha^v + \alpha^s|^2,\nonumber\\
 \Sigma^D_\phi
 &\simeq&1-2|\alpha^s|^2\frac{r_u^2}{2+r_0^2}\simeq1-0.67|\,\alpha_s|^2
 \label{A10}
 \end{eqnarray}
 where $r^2_u=(S^U_1/S^N_1)^2\simeq 1.34$ and $r^2_0=(S^N_0/S^N_1)^2\simeq
 2$;
 $\alpha^{v},(\alpha^{s})$ is the isovector
 (isoscalar) part of the unnatural parity-exchange amplitude
 relative to the dominant isoscalar natural parity-exchange amplitude
 in the $\gamma p$ reaction
 \begin{eqnarray}
 \alpha^{(v,s)}=|\frac{T_0^{U\,(v,s)}}{T_0^N}|\,{\rm e}^{i\delta^{(v,s)}},
 \label{A11}
 \end{eqnarray}
 and $\delta^{(v,s)}$ is the relative phase.
 In our case, $\alpha^{v}$ and $\alpha^{s}$ are
 identified with the strength of $\pi$ and $\eta$-exchange amplitudes
 with  $|\alpha^{s}/\alpha^{v}|\ll 1$.
 This qualitative estimation is verified
 by  the numerical calculation shown in Fig.~\ref{Fig5}a.
 The deviation of $\Sigma_\phi$ from 1 for
 $\gamma D\to \phi D$ is very small as compared to the case of
 photoproduction from the proton. On the other hand, if
 there are some non-diffraction exotic
 channels~\cite{TOYM98,Kochelev01}),
 which could generate an isoscalar unnatural parity-exchange amplitude with
 \begin{eqnarray}
 |\alpha^s|=|\alpha^{\rm exotic}|\sim |\alpha^v|,
 \label{A12}
\end{eqnarray}
 the difference $\Sigma^D_\phi-1$
 is finite and a combined analysis of
 $\Sigma^p_\phi$ and $\Sigma^D_\phi$   gives  information
 on the ratio $|\alpha^{\rm exotic}|/|\alpha^v|$ and the relative phase
 of $\alpha^s$ and $\alpha^v$.

  Complementary information about $\alpha^s$ may be obtained
  from the double beam target asymmetries.
  For the  circular polarized photon and polarized deuteron, we have
  three initial spin states with the total spin projection $J_z$:
  ($\rightrightarrows$; $J_z=2$) when the deuteron is polarized along the beam
  polarization,
  ($\leftrightarrows$; $J_z=0$) when the deuteron is polarized along the opposite
  direction  to the beam
  polarization, and
 ($\rightarrow\uparrow $; $J_z=1$) when the deuteron is polarized perpendicular   to
  the beam polarization.
  In contrast to the $\gamma p$ reaction with only one beam target
  asymmetry between $J_z=\frac32$ and $J_z=\frac12$ total spin states,
  we have three
  beam-target asymmetries
 \begin{eqnarray}
  C^{21}_{BT}&=& \frac{d\sigma(\rightrightarrows) - d\sigma(\rightarrow\uparrow)}
 {d\sigma(\rightrightarrows) + d\sigma(\rightarrow\uparrow)}\nonumber\\
  C^{20}_{BT}&=& \frac{d\sigma(\rightrightarrows) - d\sigma(\leftrightarrows)}
 {d\sigma(\rightrightarrows) + d\sigma(\leftrightarrows)}\nonumber\\
  C^{10}_{BT}&=& \frac{d\sigma(\rightarrow\uparrow) - d\sigma(\leftrightarrows)}
 {d\sigma(\rightarrow\uparrow) + d\sigma(\leftrightarrows)}
  \label{A13}
 \end{eqnarray}
 Since the amplitude of photoproduction from deuteron with a fixed spin state
 depends on the corresponding form factor
 $S$, and since these form factors are
 different for the natural and unnatural parity-exchange amplitudes, we
 expect (i) a strong influence of the unnatural parity exotic component and
 (ii) a difference in behaviour and magnitude of the various
  beam-target asymmetries.
    Using  Eq.~\ref{A5}  we can estimate the beam target
    asymmetries at $t\sim t_{\rm max}$ as follows
\begin{eqnarray}
   C^{21}_{BT}&\simeq&  \frac{1-r_0^2}{1+r_0^2} +
    2|\alpha^s|\xi\,\frac{r_u}{1+r_0^2}
   \simeq -0.33 +0.77\, |\alpha^s|\,\xi, \nonumber\\
   C^{20}_{BT}&\simeq& 2|\alpha^s|\,\xi, \qquad
   C^{10}_{BT}= - C^{21}_{BT},
  \label{A14_}
\end{eqnarray}
 where $\xi=\cos\delta^s$.
If $|\alpha^s|=|\alpha^\eta|\ll |\alpha^\pi|$ one gets the
different threshold behaviour of these two asymmetries
\begin{eqnarray}
 C^{21}_{BT}(t_{\rm max})\simeq -0.33,\qquad  C^{20}_{BT}(t_{\rm
 max})\simeq 0.
  \label{A15_}
\end{eqnarray}
 Our estimation agrees with the corresponding numerical calculation
 of $C_{BT}^{21}$ and $C_{BT}^{20}$, shown in Fig.~{\ref{Fig5}}b,
 where $\alpha_s$ is defined by the corresponding pseudoscalar and Pomeron exchange
 amplitudes described in~\cite{OTL01}.
 This result can be considered as
 a  "non-exotic" background for the $\gamma D\to \phi D$ reaction.
 If any deviation from the predicted values are measured,
 the existence of an exotic
 isoscalar unnatural parity-exchange component will be confirmed.
 For the proton target, the corresponding estimation
 reads
\begin{eqnarray}
 C^{\gamma p}_{BT}(t_{\rm max})&\simeq&
 2 (\alpha^v\cos\delta^v + \alpha^s\cos\delta^s).
  \label{A16}
\end{eqnarray}
 Therefore, the combined study of photoproduction from proton and
 deuteron can unambiguously fix the  amplitudes of the exotic channels
 and also of the isovector channel.

 The total cross section of the $\gamma D\to \omega D$ reaction is
 predicted to strongly decrease as compared to
 the $\gamma p\to \omega p$ reaction
 because the $\pi$-exchange
 amplitude, dominant in the nucleon case, is strictly suppressed for the deuteron.
 At $E_\gamma\sim 2$ GeV, the dominant channels  would
 be the Pomeron exchange and the resonance excitations.
  \begin{figure}[h]
 \includegraphics[width=0.45\columnwidth]{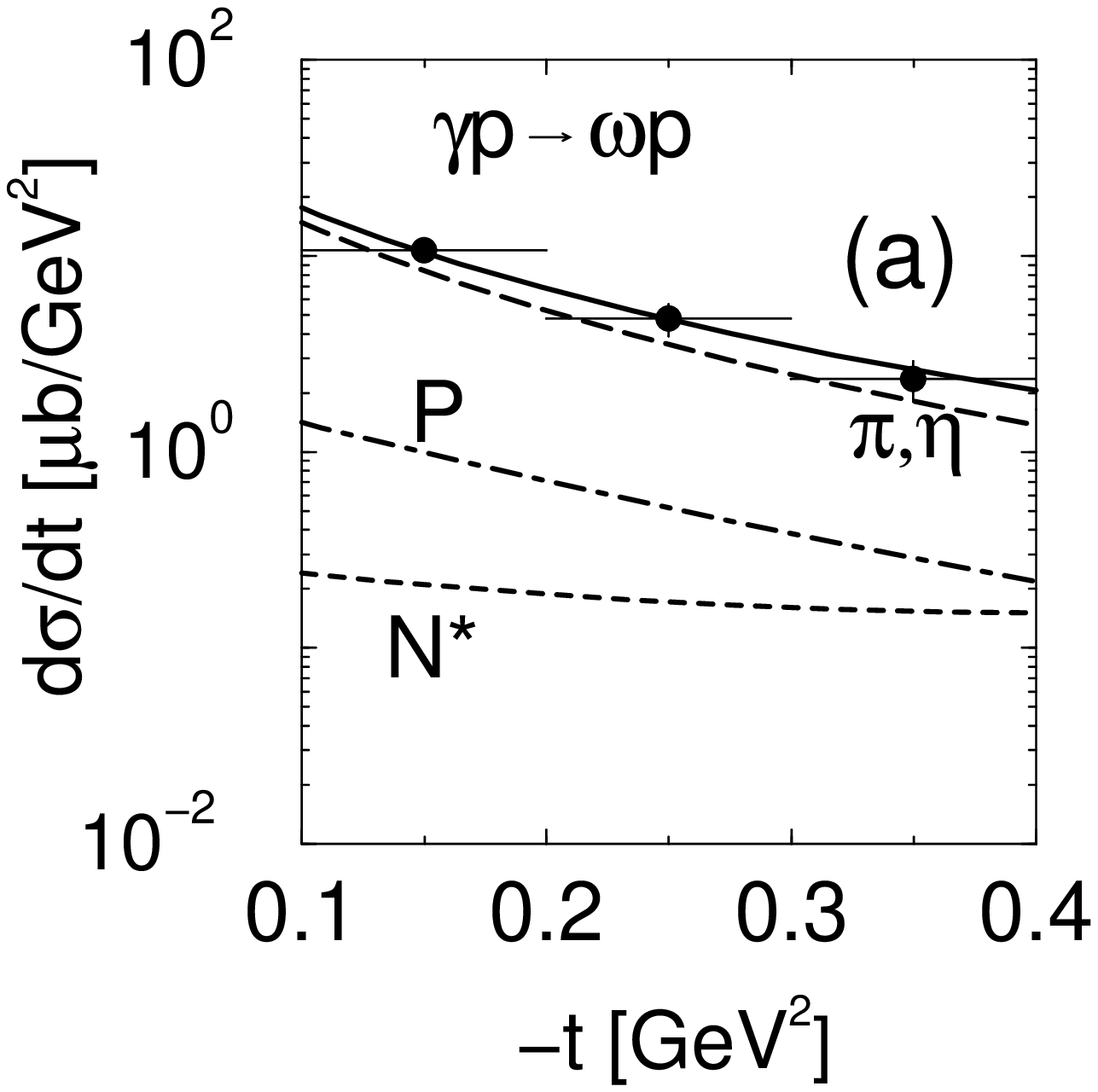}
 \includegraphics[width=0.45\columnwidth]{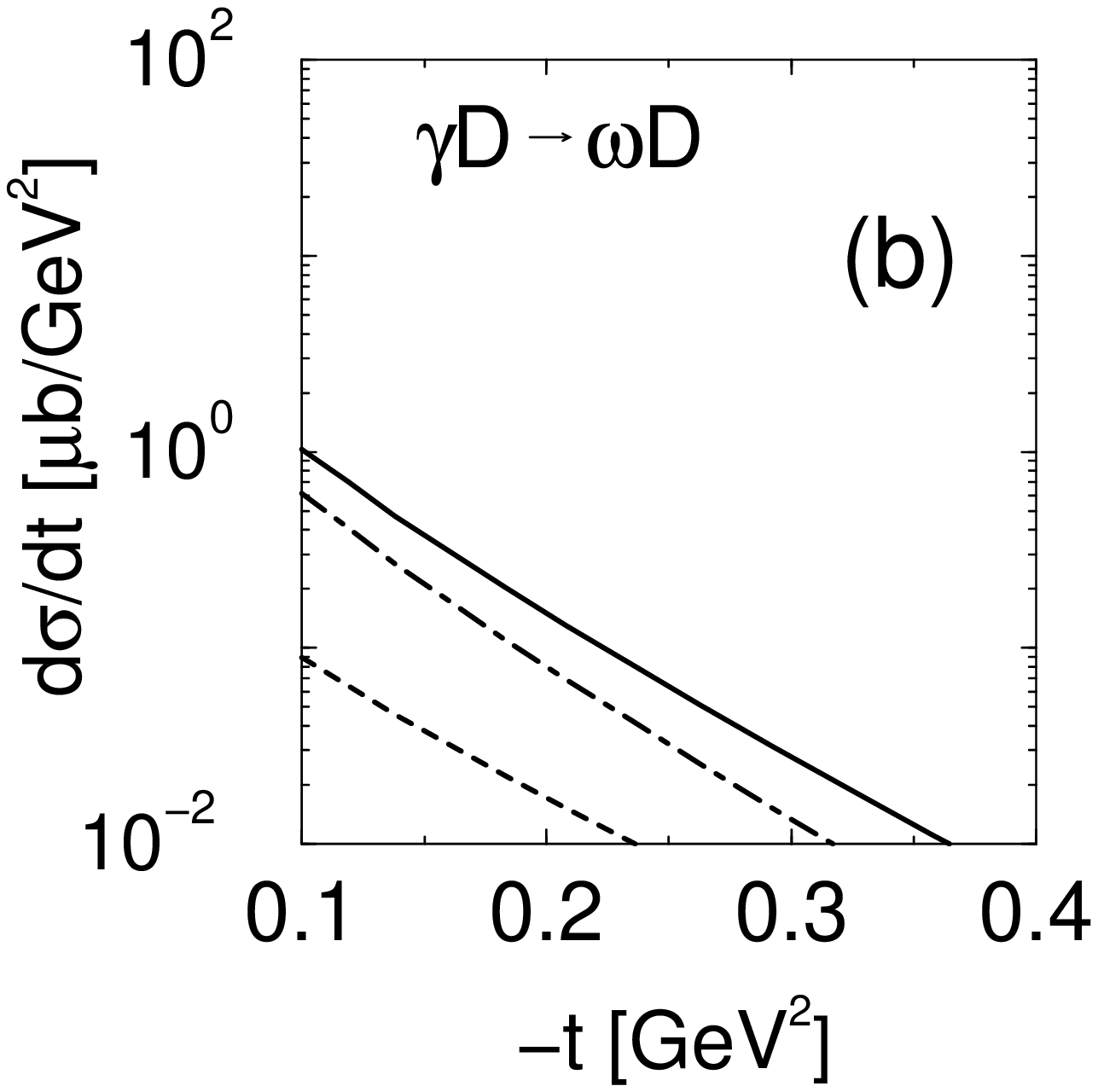}
 \caption{Differential cross sections for the $\gamma p\to \omega p$
 (a) and $\gamma D\to \omega D$ (b) reactions
 at $E_\gamma$=1.92 GeV. The contribution of $N$-channel is much smaller
 than from $N^*$ and therefore is not shown here. Data are taken from Ref.
 \protect\cite{Klein96-98}}
 \label{Fig6}
 \end{figure}
  Therefore,  we expect that the cross section
  of the $\gamma D\to \omega D$ reaction
  is of the same order of magnitude as of the  $\gamma D\to\phi D $
  reaction. In the calculation of the nucleon and the baryon resonances
  excitation channels,  we have to
  use the isoscalar coupling constants
  $g_{sNN}=(g_{pNN}+ g_{nNN})/2$ and
  $g_{sNN^*}=(g_{pNN^*}+ g_{nNN^*})/2$,  for the $\gamma NN$ and
  $\gamma NN^*$ interactions, respectively. In the present calculation, we
  include all the low-energy resonances, listed in the Particle Data Group
  using the effective-Lagrangian approach of Ref.~\cite{TL02}

\begin{figure}[h]
 \includegraphics[width=0.43\columnwidth]{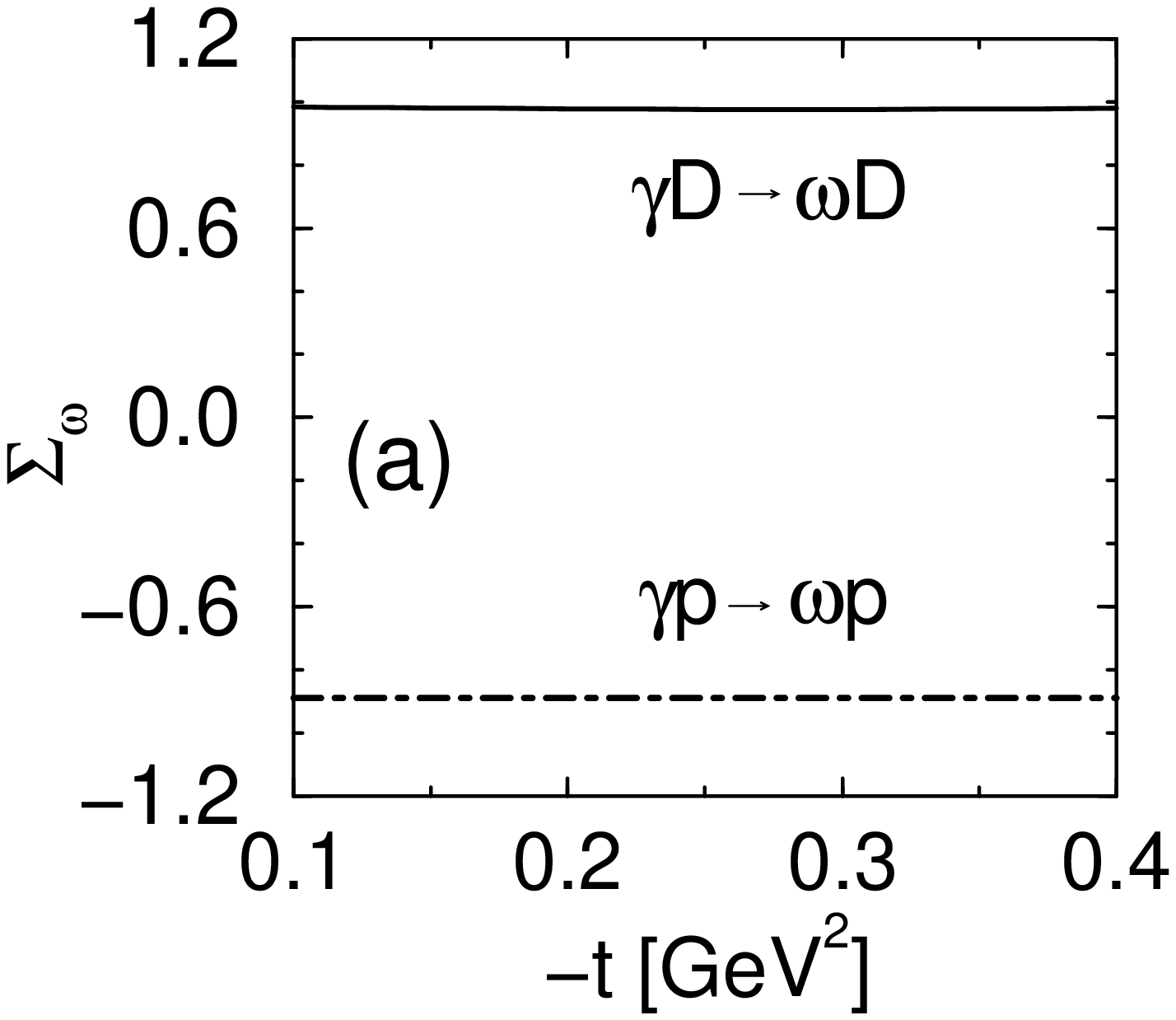}\qquad
 \includegraphics[width=0.43\columnwidth]{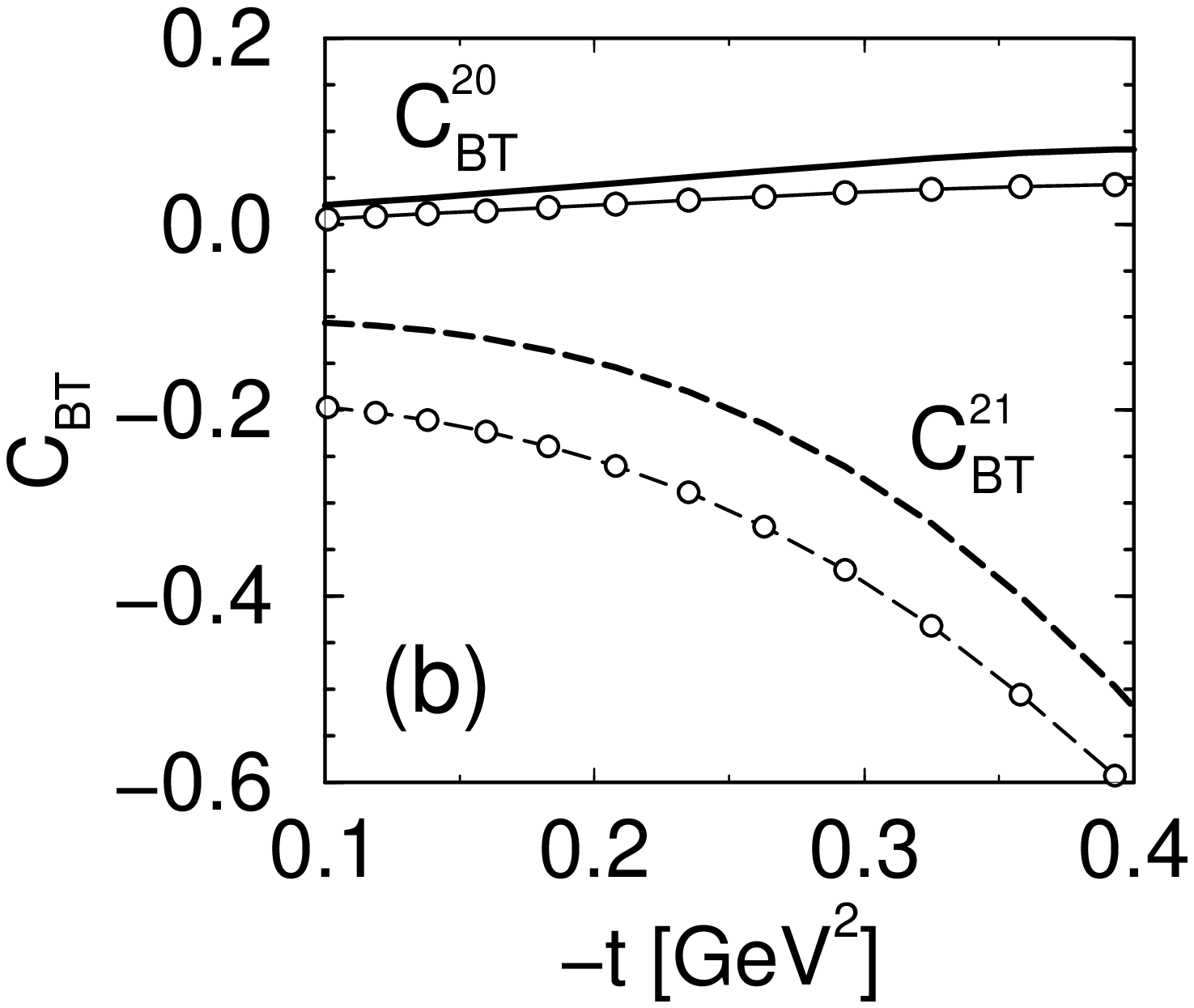}
 \caption{
 (a): The $\omega$ decay asymmetry for reactions $\gamma p\to \omega p$
 (dot-dashed line) and $\gamma D\to \omega D$ (solid line).
  The results without including $N^*$ are only 2-3\% difference
  in both cases and therefore are not shown here.
 (b): Beam-target asymmetries $ C^{21}_{BT}$ and $C^{20}_{BT}$
  for the $\gamma D\to \omega D$ reaction.
   $E_\gamma$=1.92 GeV. The open circles show the results of calculation
   without  $N^*$-channels.}
 \label{Fig7}
 \end{figure}

  Fig.~\ref{Fig6} shows the result for the
  $\gamma  p\to \omega p$ and  $\gamma D\to\omega D$ reactions at
  $E_\gamma=1.92$ GeV. The cross section of the
  $\gamma D\to\omega D$ reaction is strongly suppressed
  and becomes  comparable with
  the cross section of  $\phi$ meson photoproduction. We also see
  the dominance of the Pomeron
  exchange and strong suppression of the relative contribution
  of the unnatural parity-exchange part which stems from
  the $\eta$-exchange diagrams. This modification affects
  the $\omega$ decay asymmetry, which is shown in
  Fig.~\ref{Fig7}a.
  The asymmetry changes drastically from a value of $-0.9$
  for the $\gamma p$ reaction to +1 for the $\gamma D$ reaction
  at forward-angle photoproduction.
  Deviation from +1 for the $\gamma D\to \omega D$ reactions  would be
  a strong indication for existence of an unnatural parity-exchange
  exotic non-diffractive channels.
  Effects of  $N^*$-excitation  are only 2-3\%  and therefore are not shown in
  in  Fig.~\ref{Fig7}a.
 Our prediction for beam-target asymmetries shown in
 Fig.~\ref{Fig7}b
 is qualitatively similar to the $\phi$ meson photoproduction.
 Quantitative differences between the $\phi$ and $\omega$
 meson photoproduction in $C_{BT}^{21}$
 are understood  in part by some difference in  kimenatics
 (transverse part of ${\bf q}$ in Eq.~\ref{A2} in $\omega$-
 production is greater),
 but  mainly by essential contribution of the resonance excitations
 in  case of  $\omega$ photoproduction.

   In summary, we have shown that the coherent $\phi$ and $\omega$-meson
   photoproduction from the deuteron opens an unique
   opportunity to study the non-diffractive mechanisms with
   unnatural parity exchange exchanges,
   such as the $s\bar s$-knockout,  anomalous Regge
   trajectories,  and the  spin-flip
   excitations of baryon resonances.
   For future experimental tests, we have presented
   predictions for various spin observables in Figs. 5 and 7.

   Finally, we stress that the present investigation is a very first
   step. It would be important  to study whether the
   meson-exchange currents and relativistic effects,
   which are known to be important
   in the processes with high-momentum transfers~\cite{FKS97,Karmanov},
   are significant in the considered kinematical region.

\acknowledgments

 We thanks S.~Date, H.~Ejiri, A.~Hosaka, L.P. Kaptari,
 V.A. Karmanov, T.~Mibe and
 R.G.T.~Zegers for fruitful discussion. Especially, we thank
 T.~Hotta and T.~Nakano for calling our attention to this
 problem. This work was supported in part
 by the Japan  Society for the Promotion of Science (JSPS), and
 by U.S. Department of
 Energy, Nuclear Division, Contract N$^o$ W-31-109-ENG-38.


 \end{document}